\documentclass[aps,prc,twocolumn,preprintnumbers,amsmath,amssymb,floatfix,epsfig,A4,nofootinbib]{revtex4-2}

\usepackage{bm}
\usepackage{natbib}
\usepackage[dvips]{color}
\usepackage{url}
\usepackage{graphicx}
\usepackage{amsmath}
\usepackage{float}
\usepackage{xcolor}
\usepackage{ulem}
\usepackage{kotex}

\date{\today}

\usepackage{color} 

\newcommand\redout{\bgroup\markoverwith {\textcolor{red}{\rule[0.5ex]{2pt}{0.8pt}}}\ULon}

\begin{document}

\title{Visualization of quantum interferences in heavy-ion elastic scattering}

\author{Kyoungsu Heo}
\address{Department of Physics and Origin of Matter and Evolution of Galaxies (OMEG) Institute, Soongsil University, Seoul 06978, Korea}
\author{K. Hagino}
\address{Department of Physics, Kyoto University, Kyoto 606-8502, Japan}
\date{\today}

\begin{abstract}
We investigate various interference effects in 
elastic scattering of the $\alpha + {}^{40}\text{Ca}$ system at $E_{\rm lab}=29$ MeV. To this end, we use an optical potential model 
and decompose the scattering amplitude into four components, that is, 
the near-side and the far-side components, each of which 
is further decomposed into 
the barrier-wave and the internal-wave components. 
Each component contributes distinctively to the angular distributions, revealing unique quantum interference patterns. 
We apply the Fourier transform technique to visualize these interference effects. 
By analyzing the images at specific scattering angles, we identify the positions and intensities of peaks corresponding to each interference component. 
This analysis offers insight into structural features of the angular distribution 
which are not apparent from the differential cross sections alone. 
\end{abstract}

\maketitle

\section{Introduction}

Differential cross sections, $d\sigma/d\Omega$, obtained from nuclear scattering experiments represent 
a cumulative result of all quantum mechanical interactions between a projectile and a target nuclei. However, this observable alone may not directly reveal the complex quantum interference effects that occur during the scattering process. A differential cross section is the square of a scattering amplitude, $f(\theta)$, combining all interference effects into a single measurement. Thus, to gain deeper insight into the physical mechanisms governing the scattering process, it is helpful 
to analyze the scattering amplitude itself and disentangle the underlying interference patterns, even though the scattering 
amplitude is not an observable. 

For collisions at energies well below the Coulomb barrier, the scattering amplitude is dominated by the 
Coulomb amplitude, $f_C(\theta)$. 
In collisions of identical spin-0 boson particles, 
such as \(^{16}\text{O} + ^{16}\text{O}\) and $^{12}$C+$^{12}$C, 
the scattering amplitude can then be described using the Coulomb scattering amplitude as \( f(\theta)=
f_C(\theta) + f_C(\pi - \theta) \) \cite{bromley1961elastic,HAGINO2024138326} 
after taking into account the symmetrization of the wave function. 
See Ref. \cite{toubiana2017} for scattering of identical nuclei with arbitrary spin. 
The resultant cross sections become symmetric with respect to 
$\theta=\pi/2$ and show characteristic oscillations due to the interferences between $f_C(\theta)$ and 
$f_C(\pi - \theta)$. 

As the incident energy increases above the Coulomb barrier, nuclear interactions become more significant, absorbing a part of the elastic 
flux, and thereby the angular distributions reveal a more complex pattern  \cite{PhysRevC.20.655,SUGIYAMA199335}. 
These effects are particularly pronounced in light heavy-ion systems. 
The \( \alpha \) + \(^{40}\text{Ca}\) system is 
of particular interest in this regard, as light projectiles like \( \alpha \)-particles exhibit strong nuclear binding, 
leading to distinctive nuclear rainbow patterns in elastic scattering \cite{Khoa07}. 
This is a system with a weak absorption, and elastic scattering cross sections tend to increase 
at large angles, producing refractive effects akin to optical rainbows. 
This occurs when a part of the elastic flux is refracted by the nuclear interaction 
between the \( \alpha \)-particle and the target nucleus. In addition, cross sections often 
show anomalously large enhancements at backward angles, 
likely due to a weak absorption \cite{Planeta79}. This is referred to as anomalous large-angle scattering (ALAS). 

The nuclear rainbow effects can be interpreted as a consequence of interferences between the near-side and the far-side components of scattering amplitudes \cite{Hussein84,Fuller75}. Here, near-side scattering is associated with positive scattering angles 
dominated by 
a repulsive Coulomb interaction, while far-side scattering corresponds to negative angles influenced by an 
attractive nuclear potential. The interference between these two components leads to characteristic 
oscillatory patterns in 
differential cross sections, particularly at small and intermediate angles. 
By isolating these interference effects through a nearside-farside decomposition, one can gain a clearer understanding 
of the underlying physical mechanisms of oscillations in angular distributions\cite{Hussein84}. 

On the other hand, the ALAS can be interpreted as a consequence of interferences between 
a barrier-wave and an internal-wave \cite{Brink77}. Here, 
a barrier-wave corresponds to a reflected wave at the outer turning point of a barrier, while 
an internal-wave corresponds to a wave which penetrates the Coulomb barrier and reflects 
at the innermost turning point. 
This was initially proposed by Brink and Takigawa \cite{Brink77} and later refined for computational efficiency in Ref. \cite{Albinski82}. This approach is particularly effective when the nuclear potential features a potential pocket, allowing for distinct turning points that facilitate component separation in the scattering process. 
Interferences between the barrier wave and the internal wave yield complex oscillatory patterns 
in angular distributions, which are particularly prominent in weakly absorbing systems, such 
as $\alpha$+$^{40}$Ca. 
The barrier-wave-internal-wave decomposition is useful in examining scattering processes where a 
projectile retains transparency, as is seen in scattering systems involving a composite particle such as an \( \alpha \)-particles.

Recently, it was proposed to take a Fourier transform of scattering amplitudes 
and visualize scattering process \cite{hashimoto2023}. 
This method was applied to the nearside-farside interferences in $^{16}$O+$^{16}$O elastic scattering \cite{HAGINO2024138326}, 
providing an intuitive interpretation of interference effects by creating images of scattering 
centers on a virtual screen. 
In this way, this approach provides a deeper understanding of how scattering patterns 
evolve with incident energies and how the interference between different components affects the 
overall scattering process. That is, this method offers a novel perspective on the complex nuclear 
interactions at play and is expected to shed light on the fundamental structure of nuclear interactions.

In this paper, we apply the same technique to $\alpha$+$^{40}$Ca scattering. 
To this end, 
we decompose the scattering amplitudes into four components: 
the nearside-barrier-wave, the nearside-internal-wave, 
the far-side-barrier-wave, and the far-side-internal-wave components. 
Such combined decomposition between the nearside-farside and the barrier-wave-internal-wave components was first introduced 
in Ref. \cite{michel2001} for the $^{16}$O+$^{16}$O system. 
Each component distinctly contributes to the scattering process, resulting in complex interference 
patterns in the angular distribution. 
We shall discuss how these four components are visualized by the Fourier transform technique. 

The paper is organized as follows. In Sec. II, we analyze the angular distribution of the $\alpha$+$^{40}$Ca scattering 
at $E_{\rm lab}=29$ MeV using an optical model. We will decompose the scattering amplitudes into the four components, 
that is, the 
nearside-barrier, the nearside-internal, the farside-barrier, and the farside-internal components. 
In Sec. III, we introduce the imaging technique and analyze the scattering amplitudes by taking the Fourier transform. 
We will discuss the origin of each peak in the image by taking the Fourier transform of each component of the scattering amplitudes. 
We will then summarize the paper in Sec. IV. 

\section{Decomposition of scattering amplitudes}

We consider the elastic scattering of the $\alpha$ + $^{40}$Ca system at the incident energy of 29 MeV as 
an example. To this end, we employ the potential B in Ref. \cite{Delbar78} given by,
\begin{eqnarray}
    V(r) - iW(r) &=& U f^2(r, d_{1}, b_{1}) - i \Big[ W_{v} d f^2(r, d_{2}, b_{2}) \nonumber \\
    && - 4 b_{3} W_{D} \frac{d}{dr} f^2(r, d_{3}, b_{3}) \Big]
\end{eqnarray}
with the function \( f(r, d_{i}, b_{i}) \) given by:
\begin{equation}
    f(r, d_{i}, b_{i}) = \left(1 + \exp{\left(\frac{r - d_{i} A_{T}^{1/3}}{b_{i}}\right)}\right)^{-1},
\end{equation}
\( A_{T} \) being the mass number of the target nucleus. 
We use the depth parameters of 
\( U = 173.14 \, \text{MeV} \), \( W_{v} = 6.67 \, \text{MeV} \), and \( W_{D} = 46.92 \, \text{MeV} \), while 
the geometric parameters of \( d_{1} = 1.41 \, \text{fm} \), \( b_{1} = 1.24 \, \text{fm} \), \( d_{2} = 1.00 \, \text{fm} \), \( d_{3} = 0.62 \, \text{fm} \), and \( b_{3} = 1.04 \, \text{fm} \).

\begin{figure}
    \begin{tabular}{c}
    \includegraphics[width=1.0\linewidth]{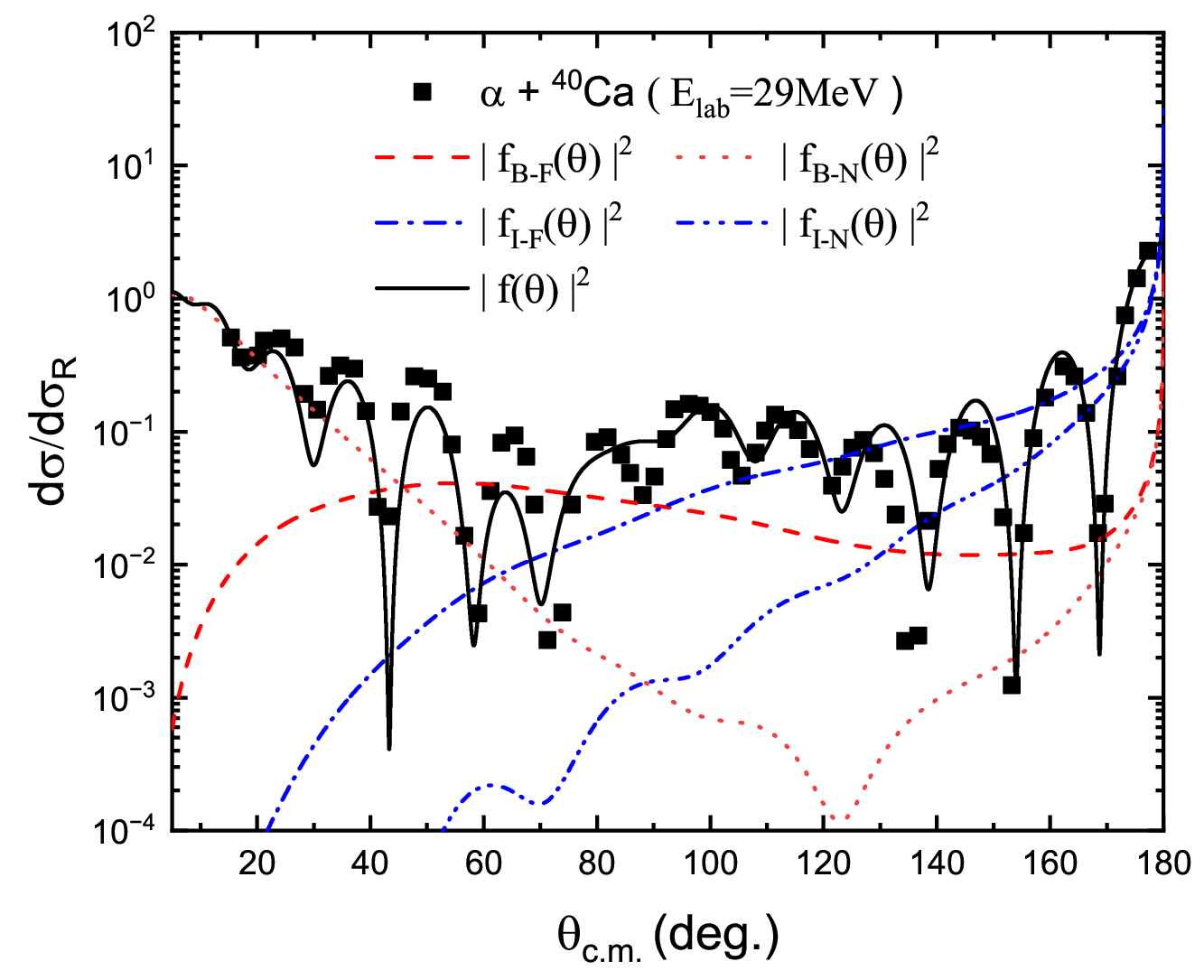}
    \end{tabular}
    \caption{The angular distribution of elastic scattering for the \( \alpha + {}^{40}\text{Ca} \) system at \( E_{\text{lab}} = 29 \, \text{MeV} \). The black squares represent the experimental data taken from Ref. \cite{Delbar78}. 
    The black solid line shows the angular distribution obtained with the 
    optical model calculation. 
    The red dashed, the red dotted, the blue dot-dashed, and the blue 
    dot-dot-dashed lines denote the angular distributions for the 
    barrier-far wave, the barrier-near wave, the internal-far 
    wave, and the internal-near wave components, respectively.}
    \label{barrierinternalnearfar}
\end{figure}

The differential cross sections for the energy 
$E=k^2\hbar^2/2\mu$, $k$ and $\mu$ being the wave number and the 
reduced mass, respectively, are given by 
\begin{equation}
    \frac{d\sigma}{d\Omega}=|f(\theta)|^2, 
\end{equation}
where the scattering amplitude $f(\theta)$ for a scattering angle $\theta$ is computed as 
\begin{equation}
    f(\theta) = f_C(\theta)+\sum_{l=0}^{\infty} (2l + 1) \frac{S_l-1}{2ik} P_l(\cos\theta). 
    \label{scatt} 
\end{equation}
Here, $f_C(\theta)$ is the Coulomb scattering amplitude, 
$S_l$ is the nuclear $S$-matrix for the $l$-th partial wave, 
and \( P_l(\cos\theta) \) is the 
Legendre polynomial.  
The angular distribution obtained with this potential is shown by the solid line in Fig. \ref{barrierinternalnearfar}, 
which well reproduces the experimental data. 

To clarify the oscillating patterns in the angular distribution, 
let us first decompose the scattering amplitudes into the nearside component \( f_N(\theta) \) and the farside 
component \( f_F(\theta) \), that is, $f(\theta)=f_N(\theta)+f_F(\theta)$. 
To separate the nearside and the farside contributions, we follow Ref. \cite{Fuller75} and replace the 
Legendre polynomials \( P_l(\cos\theta) \) in Eq. (\ref{scatt}) 
with complex combinations that selectively isolate these trajectories:
\begin{equation}
    P_l(\cos\theta) \rightarrow \tilde{Q}^{(\pm)}_l(\cos\theta) = \frac{1}{2} \left[ P_l(\cos\theta) \pm \frac{2i}{\pi} Q_l(\cos\theta)
    \right],
    \label{Q_l}
\end{equation}
where \( Q_l(\cos\theta) \) is the Legendre function of the second kind. Here, \( \tilde{Q}^{(-)}_l(\cos\theta) \) corresponds to the nearside scattering with positive deflection angles, while \( \tilde{Q}^{(+)}_l(\cos\theta) \) corresponds to the farside scattering with negative deflection angles. This replacement defines the nearside and the 
farside amplitudes as follows:
\begin{eqnarray}
    f_N(\theta) =  \sum_{l=0}^{\infty} (2l + 1) \frac{S_l-1}{2ik}\tilde{Q}^{(-)}_l(\cos\theta)+f_N^{(C)}(\theta),\\
    f_F(\theta) =\sum_{l=0}^{\infty} (2l + 1)  \frac{S_l-1}{2ik} \tilde{Q}^{(+)}_l(\cos\theta)+f_F^{(C)}(\theta), 
\end{eqnarray}    
where $f_N^{(C)}(\theta)$ and $f_F^{(C)}(\theta)$ are the nearside and the 
farside components of the Coulomb amplitude, $f_C(\theta)$ \cite{Fuller75}. 

The scattering amplitudes can be decomposed in another way, 
that is, into to the barrier wave scattering amplitude, \( f_B(\theta) \), representing reflections at the Coulomb barrier, 
and the internal wave scattering amplitude \( f_I(\theta) \), associated with reflections within 
the nuclear potential pocket. 
Those two scattering amplitudes are given as 
\begin{equation}
    f(\theta)=f_B(\theta)+f_I(\theta)
\end{equation}
with 
\begin{equation}
    f_B(\theta) = \sum_{l=0}^{\infty} (2l+1) e^{2i\sigma_l} P_l(\cos \theta) [S_B(l)-1] + f_C(\theta), 
    \label{f_B}
\end{equation}
and 
\begin{equation}
   f_I(\theta) = \sum_{l=0}^{\infty} (2l+1) e^{2i\sigma_l} P_l(\cos \theta) S_I(l), 
\label{f_I}
\end{equation}
respectively. In these equations, $\sigma_l$ is the Coulomb phase shift, while $S_B(l)$ and $S_I(l)$ are 
nuclear phase shifts for the barrier wave and the internal wave, respectively. 
Following Ref. \cite{Albinski82}, these are obtained as 
\begin{equation}
S_B=\frac{S_+S_-S_0^2}{S_++S_-^2S_0},~~~S_I=S_0-S_B,
\end{equation}
where $S_0$ is the phase shift for the original optical potential, $V(r)$, while 
$S_\pm$ are the phase shifts for modified optical potentials, $V_\pm(r)=V(r)\pm \delta V(r)$. 
We follow Ref. \cite{Albinski82} and take $\delta V(r)=\pm iW_1 e^{-(r/\rho)^4}$ with 
$W_1$ = 1.0 MeV and $\rho$ = 3.25 fm. 

By introducing $\tilde{Q}_l^{(\pm)}$ in Eq. (\ref{Q_l}) into Eqs. (\ref{f_B}) and (\ref{f_I}), 
each of the barrier and the internal wave components can further be decomposed into the near-side and the far-side 
terms\cite{michel2001}. That is, 
\begin{equation}
    f(\theta)=\sum_{X=B,I}\sum_{Y=N,F}f_{XY}(\theta), 
\label{decompositions}
\end{equation}
where \( X \in \{B, I\} \) for the barrier and the internal waves and \( Y \in \{N, F\} \) for the nearside and the farside components. 
This comprehensive decomposition highlights the role of nuclear transparency and absorption in shaping the observed 
scattering patterns. It provides a robust framework for understanding interference effects, particularly in 
systems like $\alpha$ + $^{40}$Ca at 29 MeV, where both the barrier and the 
internal wave components play a significant role.

The decomposition of the angular distributions 
is shown in Fig. \ref{barrierinternalnearfar}. 
The red dashed and the red dotted lines denote 
the angular distributions for the barrier-farside wave and the barrier-nearside wave components, respectively, while 
the blue dot-dashed and the blue 
    dot-dot-dashed lines show those for 
    the internal-farside 
    wave and the internal-nearside wave components, respectively.
One can notice that 
the nearside-barrier-wave component dominates at forward angles, 
particularly below 40°, 
where the Coulomb interaction plays a significant role. 
On the other hand, 
the two internal wave components account 
for the oscillations observed at larger scattering angles. 

\section{Visualization of Reaction processes}

\begin{figure}
    \begin{tabular}{c}
    \includegraphics[width=1.00\linewidth]{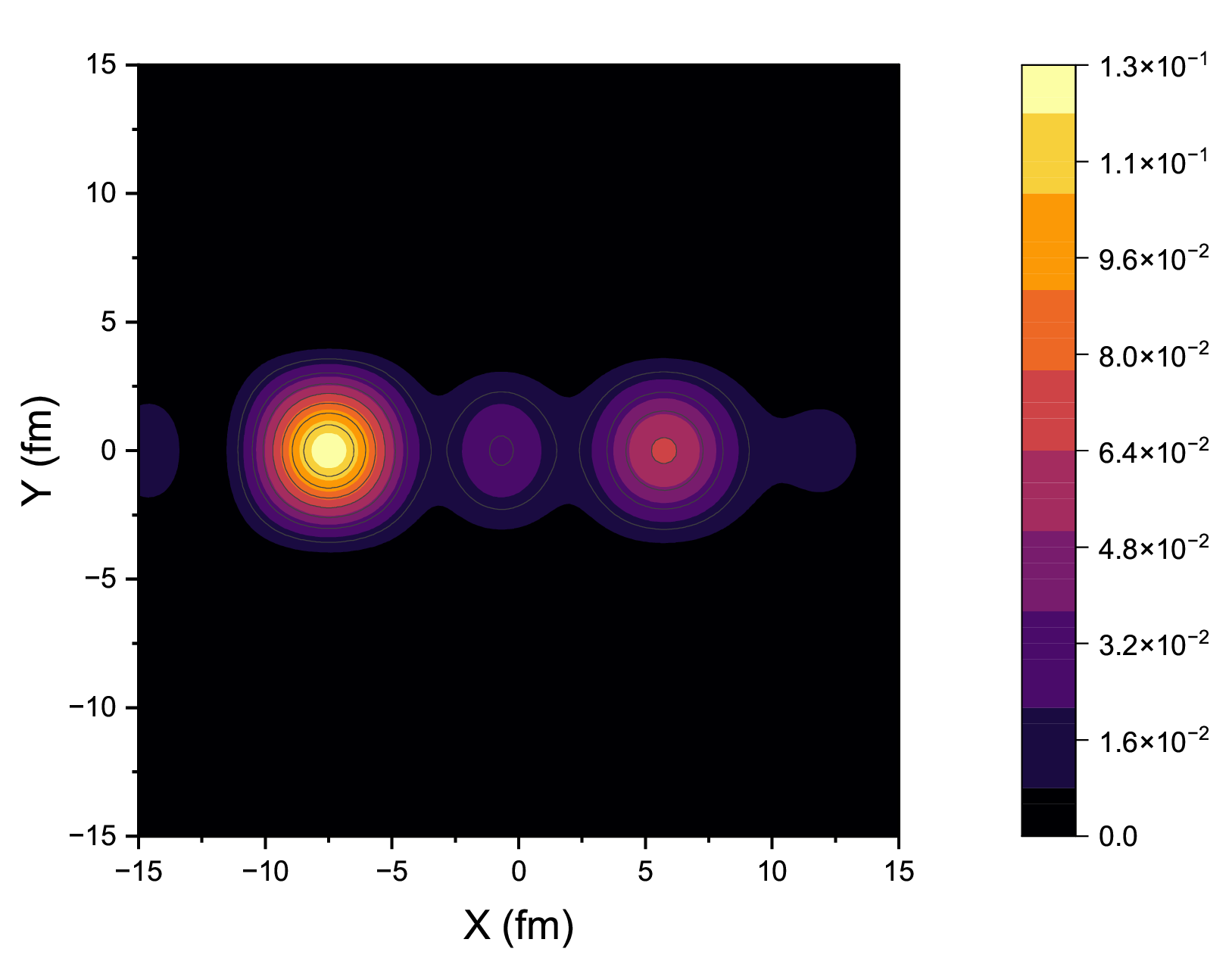}
    \end{tabular}
    \caption{The image of elastic scattering for \( \alpha + {}^{40}\text{Ca} \) at \( E_{\text{lab}} = 29 \, \text{MeV} \) and 
    $\theta$ =64°, 
    generated by Fourier transforming the total scattering amplitude. 
The angular widths are set to be \( \Delta \theta = \Delta \phi \)=15°. }
    \label{nearfarvis}
    \end{figure}

To visualize the scattering contributions and the interference effects, let us apply a Fourier transform to the scattering amplitudes\cite{HAGINO2024138326,hashimoto2023}, inspired by techniques from wave optics \cite{Hashimoto19}. This approach transforms angular 
scattering information of the scattering processes 
into a spatial image, enabling one to distinguish the interference effects from different scattering paths, 
such as the nearside-farside and the barrier-wave-internal-wave components.
In the imaging process, one introduces a lens to condense waves on a virtual screen behind the lens. 
This is formulated as follows:
    \begin{eqnarray}
    \Phi(X, Y) &=& \frac{1}{S} \int_{\theta_0 - \Delta \theta}^{\theta_0 + \Delta \theta} d\theta \, e^{i k (\theta - \theta_0) X} f(\theta) \nonumber \\
    &&\times\int_{\phi_0 - \Delta \phi}^{\phi_0 + \Delta \phi} d\phi \, e^{i k (\phi - \phi_0) Y}, 
    \end{eqnarray}
where \((X, Y)\) are coordinates on the virtual screen positioned behind the lens located at \((\theta_0, \phi_0)\) 
and  \( S = 4 (\Delta \theta) (\Delta \phi) \) represents the angular area of the lens. Here, \(\theta\) and \(\phi\) denote 
the scattering angles centered at \(\theta_0\) and \(\phi_0\) with widths \(\Delta \theta\) and \(\Delta \phi\), respectively.
The resulting image intensity, which visually represents the differential cross-section, is given by, 
    \begin{equation}
       I(X, Y) = |\Phi(X, Y)|^2. 
    \end{equation}
    
In cases where the scattering amplitude is independent of \(\phi\), the integral in the \(Y\)-direction is simplified as \cite{hashimoto2023}, 
    
    \begin{equation}
            \int_{\phi_0 - \Delta \phi}^{\phi_0 + \Delta \phi} d\phi \, e^{i k (\phi - \phi_0) Y} = 2 \Delta \phi \, \frac{\sin(k Y \Delta \phi)}{k Y \Delta \phi},
            \label{eq-Y}
    \end{equation}
which peaks at \( Y = 0 \) with an approximate width of \( 2 \pi / (k \Delta \phi) \), defining the resolution in the \(Y\)-direction.
      
Figure \ref{nearfarvis} shows the image for the total scattering amplitude. We choose $\theta_0=64^\circ$ as one can observe 
three distinct peaks in the image at this angle. 
The angular widths are set to be \( \Delta \theta = \Delta \phi =15^\circ \). One can observe the three primary peaks, 
two on the negative \(X\) side and one on the positive \(X\) side, each with distinct maximum intensities. 
The peaks observed on the \(-X\) side correspond to negative impact parameters, indicating an attractive potential 
due to the nuclear interaction and thus the farside component \cite{HAGINO2024138326}. 
On the other hand, the peak on the positive $X$ side corresponds to a positive impact parameter, thus the 
nearside component \cite{HAGINO2024138326}. 

\begin{figure}[htbp]
    \centering
    \includegraphics[width=0.4\textwidth]{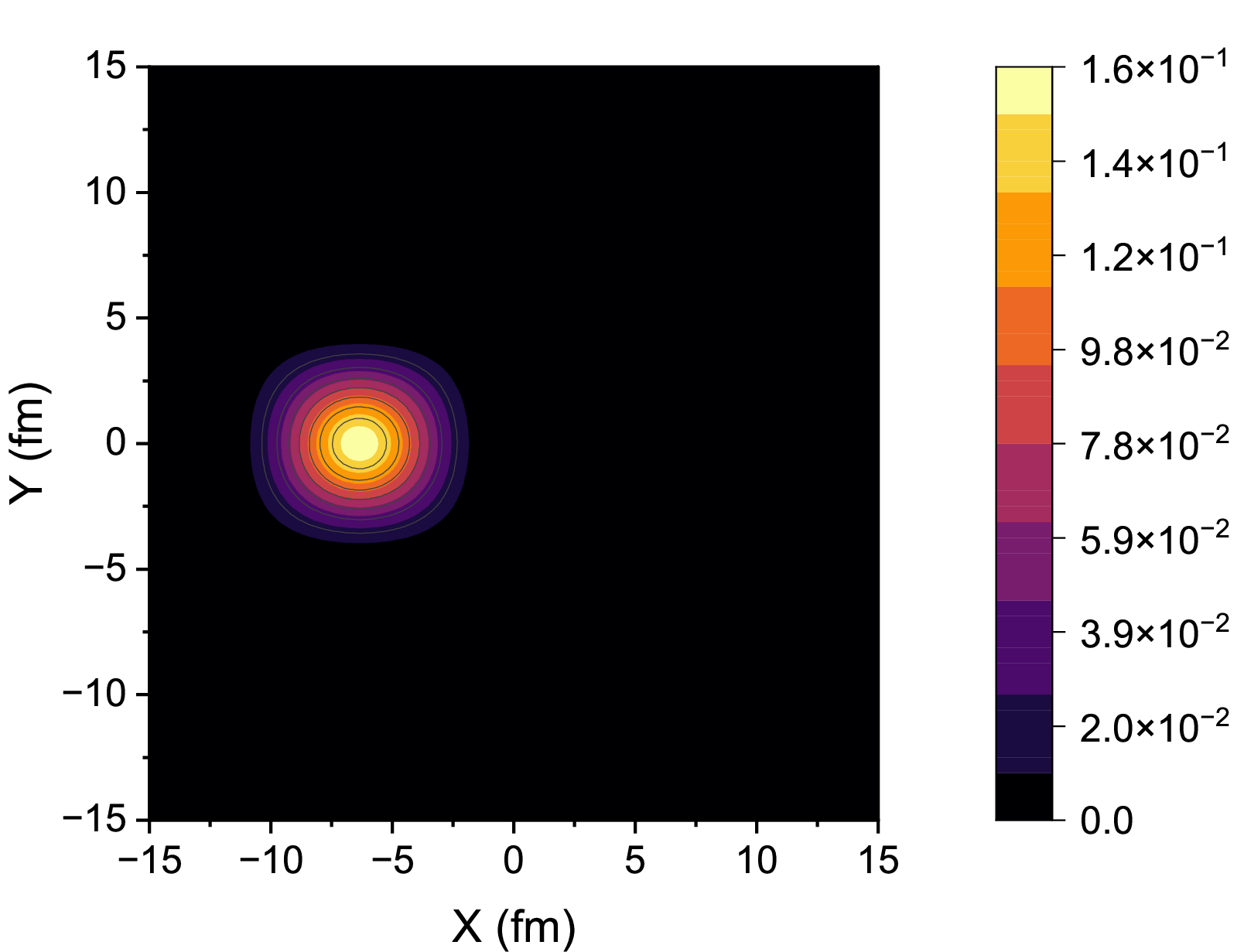}
    \hfill
    \includegraphics[width=0.4\textwidth]{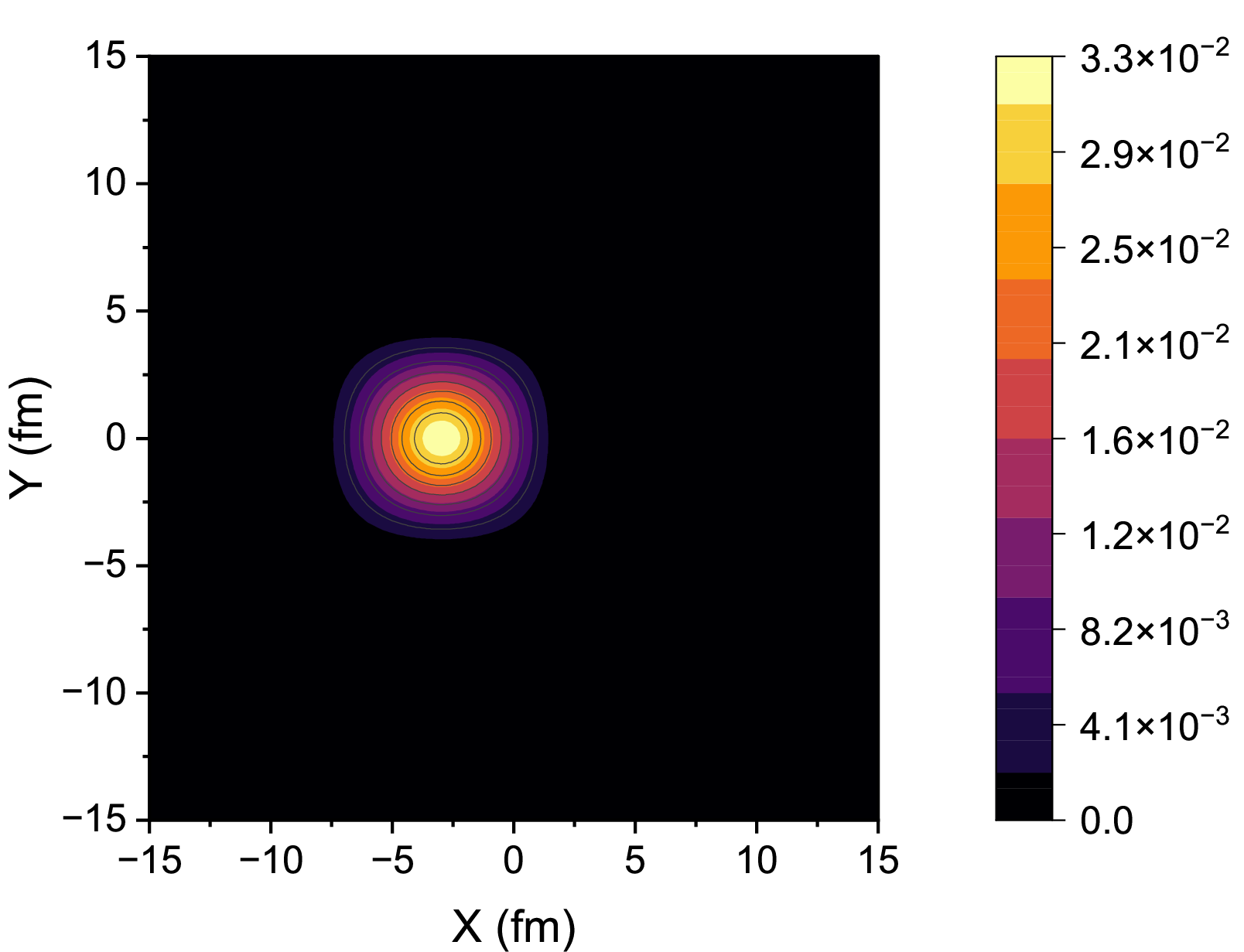}
    \hfill
    \includegraphics[width=0.4\textwidth]{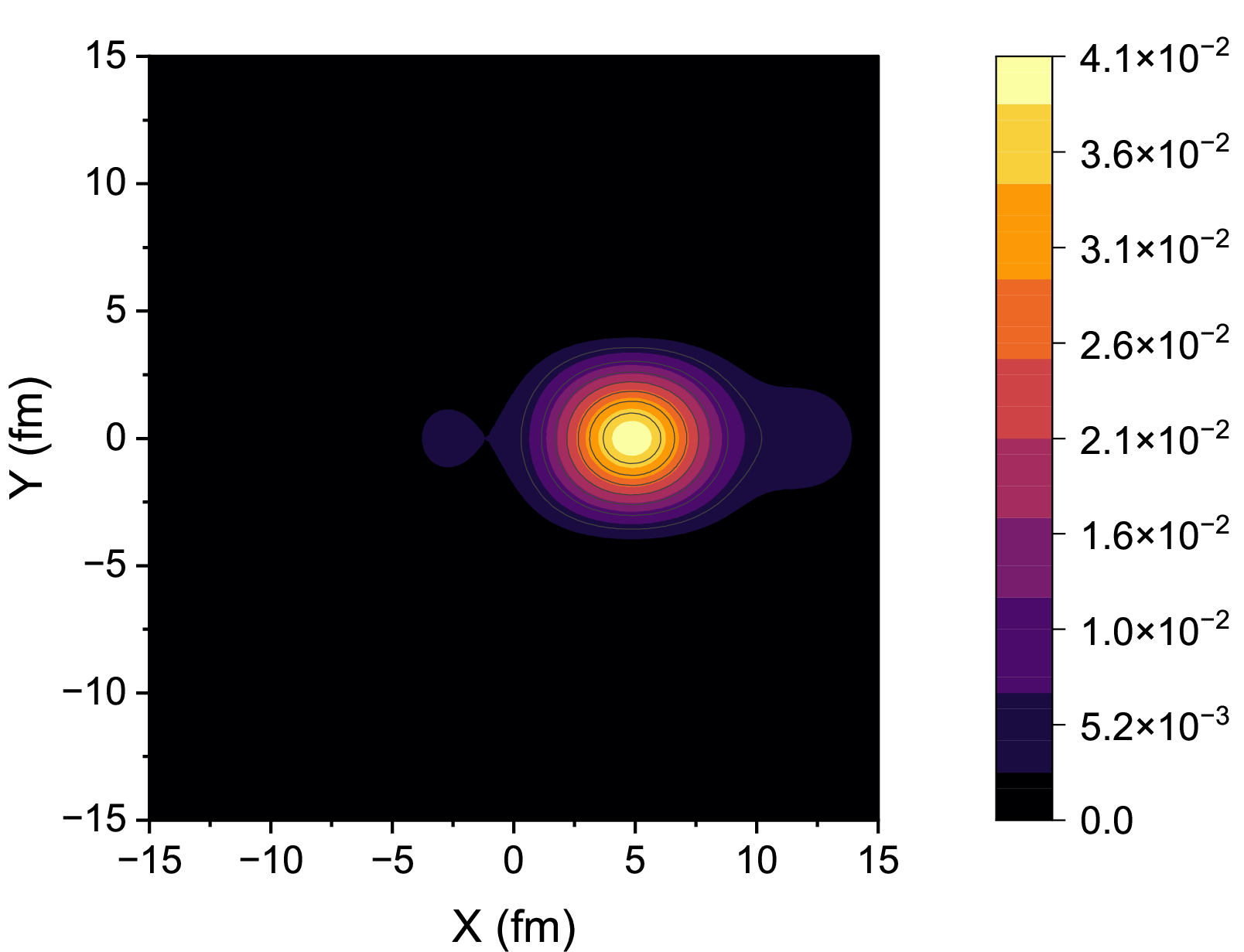}
    \caption{Same as Fig. \ref{nearfarvis}, but for the images of each component of the scattering amplitude. 
    The top, the middle, and the bottom panels show the images for 
the barrier-wave-farside, the internal-wave-farside, and the barrier-wave-nearside, respectively. }
    \label{three_imaging}
\end{figure}

In order to clarify the origin of each peak, particularly the two peaks at negative $X$, 
we decompose the scattering amplitude into the four components as in Eq. (\ref{decompositions}), and take the Fourier transform 
for each of the scattering amplitudes. 
The top, the middle, and the bottom panels in Fig. \ref{three_imaging} show the images for 
the barrier-wave-farside, the internal-wave-farside, and the barrier-wave-nearside, respectively. 
The internal-wave-nearside component has a relatively low intensity, as is indicated in Fig. \ref{barrierinternalnearfar}, 
and 
does not appear prominently in the image at this angle, even though it becomes more visible at backward angles. 
One can now see that the leftmost peak in the image shown in Fig. \ref{nearfarvis}
originates from the barrier-wave-farside component, 
and the middle peak from the internal-wave-farside component, and the peak at \(+X\) from the barrier-wave-nearside component. 
This analysis demonstrates that the imaging technique effectively captures and illustrates each interference effect in the scattering process.

\begin{figure}
    \begin{tabular}{c}
    \includegraphics[width=0.67\linewidth]{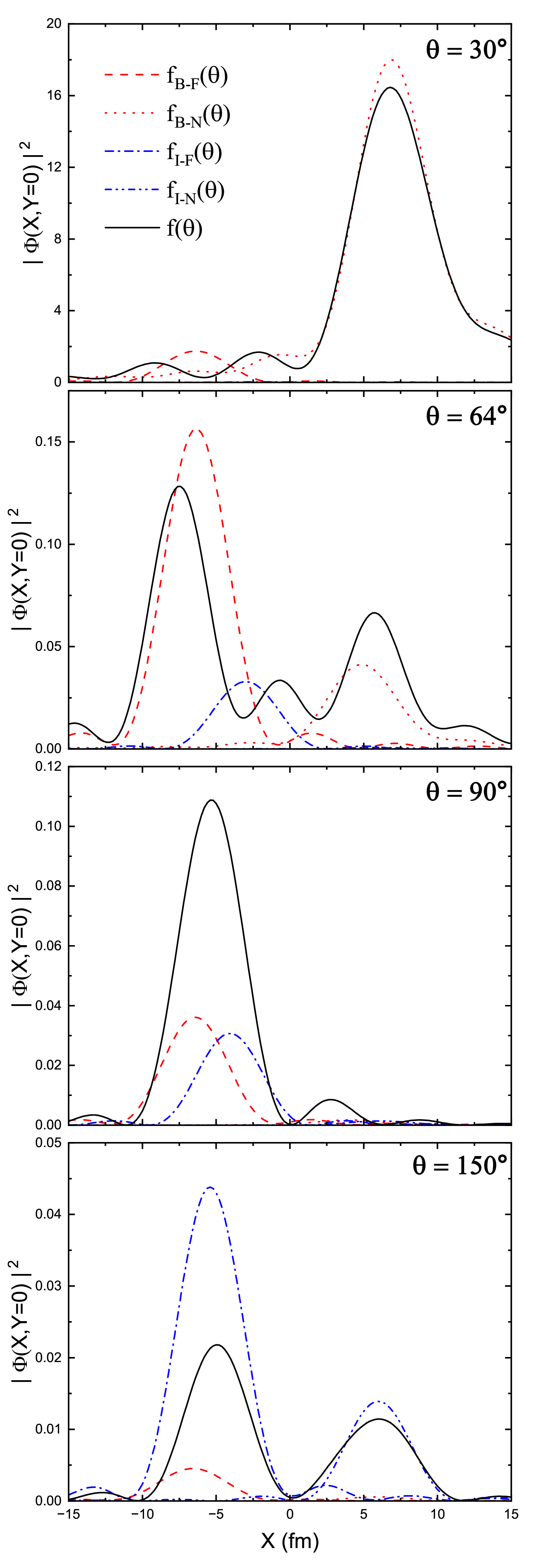}
    \end{tabular}
    \caption{Imaging results for \(Y = 0\) at four different scattering angles, \(\theta = 30^\circ\), \(64^\circ\), \(90^\circ\), and \(150^\circ\), with \(\Delta \theta = 15^\circ\). 
    The solid line represents the images from the total scattering amplitude, while the red dashed, the red dotted, 
    the blue dot-dashed, and the blue dot-dot-dashed lines denote the results for the farside-barrier wave, the 
    nearside-barrier-wave, the farside-internal wave, and the nearside-internal wave components, respectively.}
    \label{X-phi2}
    \end{figure}
Fig. \ref{X-phi2} shows the imaging results at four different scattering angles, \(\theta = 30^\circ\), \(64^\circ\), \(90^\circ\), and \(150^\circ\), all computed with \(\Delta \theta = 15^\circ\) and for \(Y = 0\). 
At \(\theta = 30^\circ\), a single prominent peak appears at \(+X\), corresponding to the 
barrier-near component. This is consistent with the fact that forward angles are dominated by the barrier wave, and near-side contributions are significant due to the relatively low deflection angles.
At \(\theta = 64^\circ\), which was also analyzed in Figs. \ref{nearfarvis} and \ref{three_imaging}, multiple peaks emerge. 
The main peak at \(-X\) is dominated by the barrier-far component, while the central peak around \(X \sim 0\) is due to both 
the internal-far and the barrier-near components. The peak at \(+X\) is associated with the barrier-near component. 
This detailed decomposition is consistent with the earlier results and confirms the roles of individual scattering amplitude 
components at this angle. Notice, however, that, due to the interference effects, the peak of each contribution does not necessarily coincide with 
the peaks for the total amplitude shown by the black solid line.
At \(\theta = 90^\circ\), the imaging result shows a single prominent peak at \(-X\), 
which arises from the barrier-far and internal-far components. The near-side contributions are suppressed 
at this intermediate angle, highlighting the dominance of farside effects.
Finally, at \(\theta = 150^\circ\), two peaks are observed: one at \(-X\) and the other at \(+X\). 
The peak at \(-X\) corresponds to the internal-far component, which becomes increasingly significant at backward angles 
due to the dominance of the internal wave. The peak at \(+X\), on the other hand, is associated with the internal-near component. 
This result highlights the growing contribution of the internal wave at large scattering angles, which is consistent with the increased 
cross sections observed at backward angles.

Overall, the imaging results clearly demonstrate how the contributions 
of each component of the scattering amplitudes evolve with the scattering angles. By representing the scattering amplitude in terms of \(X\), which is closely related to the impact parameter \cite{HAGINO2024138326}, the distinction between the far-side and the near-side components becomes immediately apparent. 
This approach reinforces the utility of imaging techniques for disentangling complex interference patterns in nuclear scattering.
Notice that the imaging technique does not require explicit decompositions of the 
total scattering amplitude to see 
how many components are contributing to the scattering at a given scattering angle.
    
\begin{figure}
    \begin{tabular}{c}
    \includegraphics[width=1.0\linewidth]{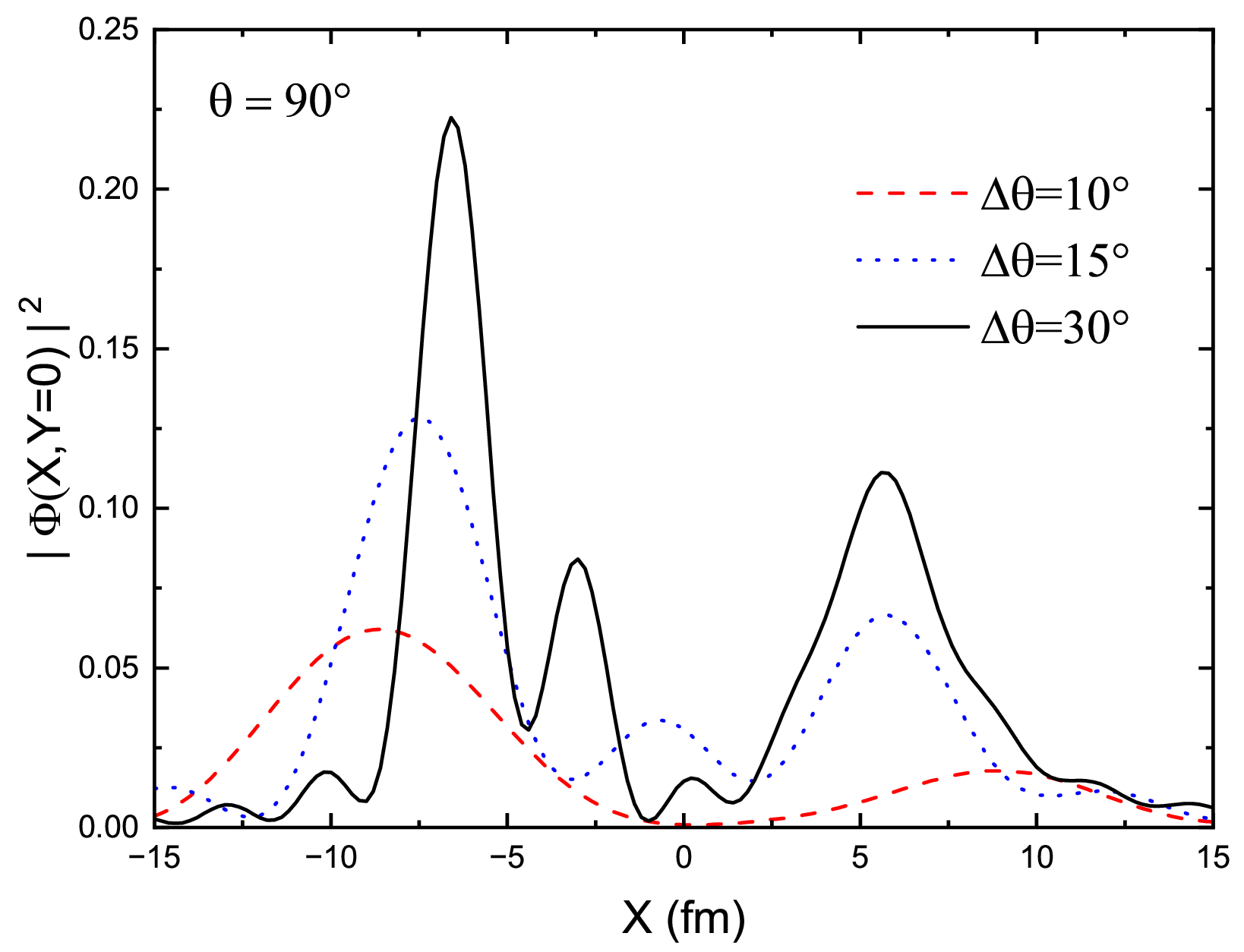}
    \end{tabular}
    \caption{The image obtained with the total scattering amplitude at $\theta=64^\circ$ with 
  three different values of $\Delta\theta$.   
} 
        \label{three_delta}
    \end{figure}
    
We notice that the choice of \(\Delta \theta\) significantly affects the resolution and interpretability of imaging results, as is indicated in Eq. (\ref{eq-Y}). Fig. \ref{three_delta} shows the imaging results for \(\theta = 64^\circ\) and \(Y = 0\) with \(\Delta \theta = 10^\circ\), \(15^\circ\), and \(30^\circ\). These results, obtained with the total scattering amplitude, demonstrate the trade-offs involved in selecting \(\Delta \theta\).
For \(\Delta \theta = 10^\circ\) (the red dashed line), two primary peaks are observed. Fine structure of the image, such as the peak around \(X \sim 0\), are suppressed due to the broad averaging effect of a small \(\Delta \theta\). 
For \(\Delta \theta = 15^\circ\) (the blue dotted line), additional structure emerges, revealing three distinct peaks. This result corresponds to the image shown in Fig. \ref{X-phi2}, confirming the consistent appearance of interference effects. 
For \(\Delta \theta = 30^\circ\) (the black solid line), more peaks are resolved as individual contributions from overlapping components become separated, leading to finer interference patterns.
This comparison highlights the balance between resolution and clarity. Smaller values of \(\Delta \theta\) emphasize 
the dominant contributions but may obscure the detailed structure of overlapping components. Conversely, larger values of \(\Delta \theta\) provide higher resolution 
but may sometimes complicate interpretation by revealing numerous peaks. 
At \(\theta = 64^\circ\), for instance, \(\Delta \theta = 15^\circ\) provides a practical balance. This value allows for resolving the primary peaks and observing the contributions from different components of the scattering amplitude. At the same time,  
minor peaks are invisible with this choice of $\Delta\theta$, which in this sense is more convenient than  $\Delta\theta=30^\circ$. 

\section{Conclusion}

We have investigated elastic scattering of the $\alpha + {}^{40}\text{Ca}$ system at $E_{\text{lab}} = 29$ MeV using an 
optical model. We have decomposed the scattering amplitude into the barrier-wave-nearside, the barrier-wave-farside, 
the internal-wave-nearside, and the internal-wave-farside components, and applied the Fourier transformation techniques to 
visualize them. 
In this way, we have identified the origins of prominent peaks in the image and matched them with the corresponding amplitude components. That is, the image has three prominent peaks at $\theta=64^\circ$, 
two peaks corresponding to negative impact parameters and one peak 
corresponding to a positive impact parameter. While the negative impact parameters correspond to the farside components, we 
have identified that the barrier-wave-farside corresponds to a larger negative impact parameter, compared to the internal-wave-farside component. The peak for a positive impact parameter corresponds to 
 the barrier-wave-nearside component. 

The analysis shown in this paper 
confirms that the imaging technique provides 
a powerful tool to analyze complex interference patterns in nuclear scattering in an intuitive manner. 
This method advances our understanding of nuclear reaction dynamics, emphasizing the role of quantum coherence 
and nuclear transparency in shaping observed scattering phenomena. 
Our results presented in this paper thus show a novel perspective for interpreting scattering mechanisms in light 
heavy-ion collisions. It would be an interesting future work to apply 
this approach to other nuclear reactions, such as proton elastic scattering, 
for which quantum effects are more prominent than in heavy-ion systems.

\section*{Acknowledgments}
\label{sec:orgae7f2b2}

This research was supported in part by Basic Science Research Program through the
National Research Foundation of Korea(NRF) funded by the Ministry of Education(Grant Nos. NRF-2020R1A2C3006177 and RS-2024-00460031) 
and in part by
JSPS KAKENHI Grant Number JP23K03414.

    \bibliographystyle{apsrev4-2}
    \bibliography{ref}
      
\end{document}